\documentstyle[preprint,aps]{revtex}
\begin{document}
\draft
\title{Chiral Baryon with Quantized Pions}
\author{J. A. McNeil}
\address{Physics Department, Colorado School of Mines, Golden,
Colorado 80401}
\author{C. E. Price}
\address{Department of Physics, University of Colorado, Boulder,
Colorado 80301}
\maketitle
\begin{abstract}
We study a hybrid chiral model for the nucleon based on the linear
sigma model with explicit quarks.  We solve the model using a
Fock-space configuration consisting of three quarks plus three quarks
and a pion as the ground state ansatz in place of the ``hedgehog''
ansatz. We minimize the expectation value of the chiral hamiltonian
in this ground state configuration and solve the resulting equations
for nucleon quantum numbers. We calculate the canonical set of
nucleon observables and compare with previous work.
\end{abstract}
\section{Introduction}
It is widely believed that quantum chromodynamics (QCD) is the
fundamental theory underlying the strong interaction.  Regrettably,
reliable first principle calculations of hadronic structure and
reactions based on QCD are still some time off. Nevertheless, simpler
QCD-motivated phenomenological models have been proposed which
preserve the important property of chiral symmetry. Familiar examples
include Skyrme-Witten models\cite{skyrme} and hybrid chiral-soliton
models such as that of Birse and Banerjee\cite{birse}.

The latter model, in particular, argues that spontaneous symmetry
breaking of the QCD lagrangian gives rise to an effective chiral
lagrangian of the Gell-Mann-L\'evy linear sigma form involving
explicit quark, scalar-isoscalar meson (sigma, $\sigma$), and
pseudoscalar-isovector meson (pion, $\vec\pi$) degrees of freedom.
They solve the model using the ``hedgehog'' ansatz which assumes a
configuration-space-isospin correlation for the pion field, $\vec\pi
= \hat r \pi$, and for the quarks. One major drawback to this ansatz
is that it breaks both rotational and isospin invariance (although
``grand spin'', $\vec K =\vec I +\vec J$, remains conserved)
requiring some projection onto physical states at the end.
Considerable attention has been  given to the problem of projection
in the calculation of observables\cite{cohen}.

In spite of this drawback the model is quite successful at predicting
baryon properties.  Constraining the pion mass and decay constant
with experimental values, the model contains but two additional free
parameters (which can be written in terms of the effective quark and
sigma masses), yet the model makes reasonable predictions for a host
of hadronic properties (mass, magnetic moments, sigma commutator,
pion-nucleon coupling constant ($g_{\pi NN}$), axial vector coupling
constant ($g_A$), as well as weak and electromagnetic form factors).

In an earlier work \cite{pands}, we attempted to avoid the serious
projection problems associated with the hedgehog ansatz by taking a
nuclear shell model approach. The resulting deformed solution for the
nucleon gave very accurate predictions for nucleon observables but
with strikingly different values for the two free parameters.
Unfortunately, this approach was also limited by a need for
projection of the total angular momentum and total isospin.

In this work we return to the hybrid chiral model of Birse and
Banerjee but avoid the hedgehog ansatz by treating the baryon as a
Fock-space admixture of quarks and quarks plus a quantized pion.
Other work\cite{araki} has used other assumed forms for the pion
field's  space-isospin correlation, but here we make no such
assumptions. The quarks and pion are simply constrained to couple to
nucleon quantum numbers; the symmetry of the quarks with respect to
permutation induces the space-isospin correlations in the pion field.
We minimize the chiral hamiltonian with respect to this
configuration.  The method is analogous to that used in the
relativistic positronium calculations of Darewicz et
al.\cite{darewicz} where the ground state is treated as an admixture
of $\{e^+ e^-\}$, $\{e^+ e^- \gamma\}$, and $\{\gamma\gamma\}$
(although we work in configuration space while Darewicz et al. work
in momentum space). In this first calculation, we have only included
$\{qqq\}$ and $\{qqq\pi\}$ in the ground state configuration, obvious
generalizations to larger configurations are straight forward and in
progress.

\section{Linear Sigma Model}

Following Birse and Banerjee we begin with the linear sigma model
lagrangian with explicit quark degrees of freedom (Arguments on how
such a form might be derived from QCD are given in Refs. [2 and 5]):

\begin{equation}
{\cal L} = \overline{q}\imath\gamma_{\mu}\partial^{\mu} q
+g\overline{q}(\sigma+\imath\gamma_5\vec\tau\cdot\vec\pi)q +{1\over
2}(\partial_{\mu}\sigma\partial^{\mu}
+\partial_{\mu}\vec\pi\cdot\partial^{\mu}\vec\pi)-U(\sigma,\vec\pi)
\end{equation}
where
\begin{equation}
U(\sigma,\vec\pi)={1\over 4}
\lambda^2(\sigma^2+\pi^2-\nu^2)^2-F_{\pi}m_{\pi}^2\sigma
\end{equation}
and where $F_{\pi}$ is the pion decay constant, $m_{\pi}$ is the pion
mass, and $\nu$, $g$ and $\lambda$ are constants to be determined. In
the standard scenario spontaneous symmetry breaking generates masses
for the quark and sigma fields and the linear sigma term, which
breaks the chiral symmetry and generates the effective pion mass. The
vacuum then has a unique non-vanishing scalar field expectation
value:
\begin{equation}
{\partial U\over \partial\vec\pi}=0\longrightarrow   \vec\pi_0=0,
{}~~~~~~~~~~~~{\partial U\over  \partial\sigma}=0\longrightarrow
\sigma_0=-F_{\pi}.
\end{equation}
Then the three undetermined constants in the original Lagrangian can
be written in terms of the three effective
masses: $m_q=-g\sigma_0$, $m_{\sigma}^2=\lambda^2(3
\sigma_0^2-\nu^2)$, and $m_{\pi}^2=\lambda^2(\sigma_0^2-\nu^2)$.

Expanding the scalar field about $\sigma_0$, $\sigma
=\sigma_0+\tilde\sigma$, and dropping constant terms, one has the
effective lagrangian:
\begin{equation}
{\cal L}_{eff}=\overline{q}(\imath\gamma_{\mu}\partial^{\mu}-m_q)q
+g\overline{q}(\tilde\sigma+\imath\gamma_5\vec\tau\cdot\vec\pi)q
+{1\over 2}(\partial_{\mu}\sigma\partial^{\mu}
+\partial_{\mu}\vec\pi\cdot\partial^{\mu}\vec\pi)    -{\cal
V}(\tilde\sigma,\vec\pi)
\end{equation}
with
\begin{equation}
{\cal
V}(\tilde\sigma,\vec\pi)={\lambda^2\over4}[\tilde\sigma^4+\vec\pi^4
-4\tilde\sigma^3F_{\pi}+2\tilde\sigma\vec\pi^2(\tilde\sigma-2F_{\pi})
].
\end{equation}
We take the experimental values $F_{\pi}=93$ MeV and $m_{\pi}=139.6$
MeV, leaving $m_q$ and $m_{\sigma}$ as the only free parameters which
must be determined by fitting to nucleon observables.

\section{Ground State Configuration}

To solve this model we consider the following Fock-space ground state
configuration:
\begin{equation}
|jj_z;tt_z>=\{{A}[b_1^{\dagger}b_2^{\dagger}b_3^{\dagger}]_{jj_z;tt_z
}+{B}
[b_1^{\dagger}b_2^{\dagger}b_3^{\dagger}a_4^{\dagger}]_{jj_z;tt_z}\}|
0>
\end{equation}
where $b_i^{\dagger}\equiv b^{\dagger}_{{1\over2}m_i;{1\over2}t_i}
(a_i^{\dagger}\equiv a^{\dagger}_{{1}m_i;{1}t_i})$ refer to creation
operators for quarks (pions) in the $1s{1\over2}~~(1P)$ state. Note
that both operators obey commutation (not anticommutation) relations.
\begin{equation}
[b_{\alpha}^{\dagger},b_{\beta}]=\delta_{\alpha\beta},
\end{equation}

\begin{equation}
[a_{\alpha}^{\dagger},a_{\beta}]=\delta_{\alpha\beta}
\end{equation}

\begin{equation}
[b_{\alpha}^{\dagger},a_{\beta}]=[b_{\alpha},a_{\beta}]=0.
\end{equation}
The antisymmetry of the fermions (quarks) is included via the color
degree of freedom which is not shown explicitly.

The bracket notation refers to the spin-isospin coupling, e.g.

\begin{eqnarray}
[b_1^{\dagger}b_2^{\dagger}b_3^{\dagger}]_{jj_z;tt_z}=
&{1\over \sqrt{12}}{\sum\atop{\scriptstyle j_{12},m_{12},t_{12}} }
{\sum\atop{\scriptstyle m_1,m_2,m_3,m_4\atop \scriptstyle
t_1,t_2,t_3,t_4} }<{1\over 2} m_1 {1\over 2} m_2|j_{12}m_{12}>
<j_{12}m_{12}{1\over 2} m_3|jj_z>\cr
& <{1\over 2} t_1 {1\over 2} t_2|j_{12}t_{12}><j_{12}t_{12}{1\over 2}
t_3|tt_z> b_{{1\over 2} m_1;{1\over 2} t_1}^{\dagger}b_{{1\over 2}
m_2;{1\over 2} t_2}^{\dagger}b_{{1\over 2} m_3;{1\over 2}
t_3}^{\dagger}
\end{eqnarray}
which is just the usual SU(6) wavefunction, and
\begin{eqnarray}
&[b_1^{\dagger}b_2^{\dagger}b_3^{\dagger}a_4^{\dagger}]_{jj_z;tt_z}=

{1\over \sqrt{18}}{\sum\atop{\scriptstyle
j_{12},m_{12},t_{12}\atop\scriptstyle  j_{123},m_{123}t_{123}} }
{\sum\atop{\scriptstyle m_1,m_2,m_3,m_4\atop \scriptstyle
t_1,t_2,t_3,t_4} } <{1\over 2} m_1 {1\over 2}
m_2|j_{12}m_{12}>\hfill\cr
&<j_{12}m_{12}{1\over 2} m_3|j_{123}m_{123}><j_{123}m_{123}1m_4|jj_z>
<{1\over 2} t_1 {1\over 2} t_2|j_{12}t_{12}>\hfill\cr
&<j_{12}t_{12}{1\over 2} t_3|j_{123}t_{123}><j_{123}t_{123}1t_4|tt_z>
b_{{1\over 2} m_1;{1\over 2} t_1}^{\dagger}b_{{1\over 2} m_2;{1\over
2} t_2}^{\dagger} b_{{1\over 2} m_3;{1\over 2} t_3}^{\dagger}a_{1
m_4;1 t_4}^{\dagger}~.\hfill
\end{eqnarray}
Note that the permutation symmetry with respect to the quark
operators insures that intermediate spin and isospin seniority values
are correlated. It is this symmetrization property  which insures the
correlation of the pion spatial and isospin components and therefore
the correlation need not be imposed through the hedgehog ansatz.

We minimize the expectation value of the hamiltonian (derived from
the effective lagrangian) in our ground state treating the sigma as
coherent state (classical field) and the pion as a quantum field.

To insure stability of this limited model it is necessary to treat
approximately the $\vec\pi^4$-terms which, aside from mass
renormalizations,  would otherwise vanish in our ground state. (This
approximate treatment would not be necessary if we expand our ground
state configuration to include a three quark-two pion term.) In the
spirit of the mean-field approximation we assume all quantities are
already vacuum dressed and, therefore, remove all vacuum terms by the
approximation:
\begin{equation}
:\vec\pi^4:\rightarrow  :\vec\pi^2: :\vec\pi^2:
\end{equation}
We consider the nucleon sector, $j={1\over 2},t={1\over 2}$, and
limit our quark wavefunction to a single orbital ($nlj=1s{{1\over
2}}$).  Writing \cite{bandd}
\begin{equation}
q^{nl}_{jm;tt_z}=

\pmatrix{{iG_{nlj} \over r}{\varphi}^+_{jm}\cr

              {-F_{nlj} \over r}{\varphi}^-_{jm} } {\eta}_{tt_z},
\end{equation}
and limiting our pion wavefunction to a single orbital (L=1) as well,
\begin{equation}
\pi_{\scriptscriptstyle LM}=\pi_{\scriptscriptstyle L}(r)
Y_{\scriptscriptstyle LM}(\hat r),
\end{equation}
we minimize the ground state expectation value of the hamiltonian to
obtain the following set of coupled radial equations:
\begin{equation}
-{dG\over dr}+{G\over r} + m_q F - g\tilde\sigma F - 2AB\sqrt{3} g\pi
G = -E G
\end{equation}
\begin{equation}
-{dF\over dr}-{F\over r} + m_q G - g\tilde\sigma G + 2AB\sqrt{3} g\pi
F = E F
\end{equation}

\begin{equation}
(\nabla^2-m_{\sigma}^2)\tilde\sigma-\lambda^2[-3F_{\pi}\tilde\sigma^2

+\tilde\sigma^3-{B^2\over 3}(F_{\pi}-\tilde\sigma)\pi^2]=-g\rho_s
\end{equation}

\begin{equation}
(\nabla^2-m_{\pi}^2)\pi-\lambda^2[(\tilde\sigma^2-2\tilde\sigma
F_{\pi})\pi+\pi^3)=-2\sqrt{3}{A\over B} g\rho_{ps}
\end{equation}
where
\begin{equation}
\rho_s={3\over 4\pi}{(G^2-F^2)\over r^2}~~~~~~~~~ \rho_{ps}=-{3\over
4 \pi}{2GF\over r^2}
\end{equation}
and the quantum number labels have been suppressed for clarity.
Minimization with respect to A (B) yields:
\begin{equation}
{B\over A}={{\lambda^2\over 4}\int d^3x \pi^4 \over

             \sqrt{3}g\int d^3x\rho_{ps} \pi}
\end{equation}
We solve the coupled equations self-consistently from starting
densities chosen strictly for convenience (the final solution is
insensitive to the details of the starting point).

\section{Results}

First we consider the Birse-Banerjee parameters, $m_q$=500 MeV and
$m_{\sigma}$=1200 MeV.  We obtain self-consistent solutions with a
net nucleon energy (before center of mass correction) of 1196 MeV
compared with 1116 MeV obtained using the hedgehog ansatz. The close
comparison is a bit deceiving however because the energy
contributions from the various sources, given in Table 1, differ
considerably.  In particular the pion kinetic energy is much smaller,
164 MeV in our case compared to 565 MeV from Birse and Banerjee. This
small contribution from the pion is similar to our earlier deformed
nucleon result \cite{pands}.

In Fig.\ \ref{fig1} we compare the quark densities and the scalar and
pseudoscalar fields. Clearly, our Fock-space soliton is somewhat more
compact which implies a smaller rms size for the soliton (and
explains the much higher central quark density). Surprisingly, the
Fock-space pion field is much larger that the hedgehog pion field in
apparent contradiction with the small pion contribution to our total
energy. Evidently, the shape of the pion field is almost entirely due
to its nonlinear interaction with the other fields. This is in sharp
contrast with our earlier results for the deformed soliton where both
the pion field and its contribution to the total energy were quite
small.

Once the ground state configuration is solved for a given parameter
set, one can calculate the baryon observables.  Besides the mass, we
calculate and compare with previous work the nucleon magnetic
moments, $\sigma$-commutator,  axial vector coupling ($g_A$),
pion-nucleon coupling ($g_{\pi NN}$) and charge radius. The magnetic
moment is given by

\begin{equation}
\mu=<gs|{e\over 2}(\vec r\times\vec j)\cdot\hat z|gs>
\end{equation}
with $  j_i =({1\over6}+{1\over2}\vec\tau)\gamma_i-\vec\pi\times
\nabla_i\vec\pi$ (which contains both isoscalar and isovector
pieces). The axial vector coupling parameter, $g_A$, is given by
\begin{equation}
g_A=2<gs|A_3^z|gs>
\end{equation}
\begin{equation}
A_3^z={1\over
2}\overline{q}\gamma_5\gamma^z\tau_3q-\sigma{\partial\over\partial
z}\pi_3
+\pi_3{\partial\over\partial z}\sigma.
\end{equation}

The sigma-commutator is given by
\begin{equation}
\sigma(\pi N)=F_{\pi}m_{\pi}^2<gs|\int d^3r \tilde\sigma(\vec r)|gs>.
\end{equation}

The pion-nucleon coupling constant is given by
\begin{equation}
{g_{\pi NN}\over 2 M}=<gs|z j_{\pi,3} |gs>
\end{equation}
where
\begin{equation}
j_{\pi,3}=ig\overline{q}\gamma_5\tau_3q-\lambda^2(\sigma^2+\vec\pi^2-
F_{\pi}^2)\pi_3.
\end{equation}

Table 2 gives these observables for our model in terms of the various
radial integrals involved.  Note that the pseudoscalar radial
integral ($I_1$) enters the quark contributions to both the magnetic
moment and the pion-nucleon coupling.  There is the nontrivial
question of which ``nucleon'' mass should scale the integral in each
case.  To compare magnetic moments with standard experimental values
the nucleon mass, which enters via the definition of the nuclear
magneton, must be used to scale these integrals; however, for the
pion-nucleon coupling constant the natural quantity is the ratio of
the coupling constant to the chiral baryon mass.

In Table 3 we present the values for each of the observables and
compare with Birse-Banerjee.  First the magnetic moments in the
present model are somewhat smaller than either the hedgehog case or
experiment. In particular the pion contributions are significantly
smaller (by almost a factor of 4). It is an open question whether the
pion contribution in an expanded configuration ansatz will improve
this value. The axial vector coupling is likewise smaller in our case
(1.41 versus 1.86) but closer to experiment (1.25). We find the value
1.25 ($2M/m_{\pi}$) for the  pion-nucleon coupling constant compared
to 1.53 from the hedgehog ansatz and the experimental value of 1.0.
Finally we find 36.9 MeV for the sigma commutator versus 94.0 MeV
from the hedgehog ansatz compared to a somewhat uncertain 25-60 MeV
experimental value.\cite{gasser} With the important exception of the
magnetic moments the calculated baryon observables in the present
model are quite close to the experimental values in spite of the
extremely limited Fock-space ground state. Finally, the

Goldberger-Trieman test,
\begin{equation}
g_{\pi NN}{m_{\pi}\over 2 M}\simeq g_A{ m_{\pi}\over 2 F_{\pi}},
\end{equation}
is reasonably satisfied with $1.25\approx1.05$ compared to $1.0
\approx .93$ experimentally.

Since these parameters ($m_q=500 MeV$ and $m_\sigma=1200 MeV$) were
tuned to provide the best possible hedgehog description of nucleon
observables, it is appropriate to repeat this comparison using
parameters that are chosen to optimize the Fock-space soliton
predictions. In Fig.\ \ref{fig2}, we show the quark densities and
meson fields obtained from our Fock-space soliton using $m_q=375 MeV$
and $m_\sigma=600 MeV$. The quark density is similar to that obtained
from the hedgehog soliton and would clearly lead to a larger charge
radius that we obtained using the hedgehog parameters. Likewise, the
pion and sigma fields extend to larger distances and are somewhat
reduced in size as compared to our Fock-space soliton shown in Fig.\
\ref{fig1}; however, the pion field is still significantly larger
than that of the hedgehog soliton. Notice also that the central value
of the sigma field is less than the chiral value of $2F_\pi$ perhaps
indicating that we are near the region of parameter space for which
the soliton would dissolve (i.e. where there is no bound solution for
the nucleon).

In Table IV, we show the contributions to the total energy as in
Table I. These results are qualitatively similar to those obtained
with the hedgehog parameters with the primary difference being that
the various meson interaction terms are somewhat smaller, as would be
expected based on the smaller meson fields (shown in Fig.\
\ref{fig2}). Also of interest is that the CM corrected nucleon mass
is essentially equal to the experimental value. The other calculated
observables are shown in Table V. The charge radius and magnetic
moments are in significantly better agreement with  experiment, while
the couplings are qualitatively unchanged and only the sigma
commutator is in worse agreement. Overall, our predictions of nucleon
observables are qualitatively as accurate as those of the hedgehog
soliton. This is quite surprising considering the limited nature of
our starting ansatz.

\section{Conclusions}

In this paper we have investigated the feasibility of  using a
multi-configurational Fock-space approach to solving the hybrid
chiral sigma model. We found that even with the simplest non-trivial
configuration we could obtain reasonable predictions for nucleon
observables. The primary advantage of this approach is that the
nucleon quantum numbers are imposed from the start so that no
projection is needed to access the component of the solution
corresponding to a nucleon. This constrains the field variations to
the physical sector as opposed to a hedgehog approach where the field
variations are made before  projection onto the physical sector. This
early success gives us reason to hope that a more complete starting
ansatz involving multi-pion components and higher order quark and
pion wavefunctions will be able to provide accurate descriptions for
a variety of low-lying hadrons. Such calculations are currently in
progress.

%
%
%
\section{Acknowledgements}
We thank J. R. Shepard for helpful comments and suggestions. This
work was supported in part by the Department of Energy and the
National Science Foundation.

\newpage
%
%
%

%
%
%
%
\begin{figure}
\caption{Quark densities and meson fields for the Fock-space and
Hedgehog solitons using $m_q$= 500 MeV and $m_{\sigma}$= 1200 MeV.}
\label{fig1}
\end{figure}
\begin{figure}
\caption{Quark densities and meson fields for the Fock-space soliton
using $m_q$= 375 MeV and $m_{\sigma}$= 600 MeV.}
\label{fig2}
\end{figure}

%
%
%
\begin{table}
\caption{ Energy contributions to Chiral Baryon using $m_q$=500 MeV
and $m_{\sigma}$=1200 MeV. (All values are in MeV.)  Fock-space
soliton solution yields A=0.827 and B=0.561.}
\begin{tabular}{lrr}
Quantity & This work &  Ref.[2]\\
\tableline
Quark eigenenergy                 & 123.8 &   30.5\\
Quark kinetic energy              &  1451 &   1219\\
Sigma kinetic energy               &  335 &    358\\
Pion kinetic energy               &   164 &    565\\
Quark-$\sigma$ interaction energy &  -137 &   -184\\
Quark-pion interaction energy     &  -940 &   -943\\
Meson interaction energy          &   318 &    101\\
Baryon Mass                       &  1196 &   1116\\
Center of Mass Correction         &  -303 & \\
CM corrected Baryon Mass          &   893 & \\
\end{tabular}
\end{table}

\begin{table}
\caption{ Nucleon observables for quantized pion ansatz in terms of
radial integrals. }
\begin{tabular}{ll}
Quantity  &  Form\\
\tableline
rms Charge radius & $<r^2>_{ch}^{1/2} = I_0$\\
Proton magnetic moment (Quark contribution) &$\mu_p=(A^2+{2\over
3}B^2)I_1$\\
Neutron magnetic moment  (Quark contribution) &$\mu_n=-{2\over
3}(A^2+{2\over 3}B^2)I_1$\\
Magnetic moments (Meson contribution)  &$\pm {\pi\over 3}B^2 I_2$\\
Axial vector coupling, $g_A$ (Quark contribution)  &$({5\over
3}A^2+{11\over 9}B^2)I_3$\\
Axial vector coupling, $g_A$ (Meson contribution)  &${16\pi\over
9\sqrt{3}}AB I_4$\\
Pion-nucleon coupling, $g_{\pi NN}/2M$ (Quark contribution)
&$-g({5\over 3}A^2+{11\over 9}B^2)I_1$\\
Pion-nucleon coupling, $g_{\pi NN}/2M$ (Meson contribution) &$
-\lambda^2 {8\pi\over 9\sqrt{3}}AB I_5$\\
\tableline
$I_0=\int_0^\infty dr (G^2 - F^2)$$I_1=-{2\over 3}\int_0^{\infty}dr r
GF$\\
$I_2={2\over 3}\int_0^{\infty}dr r^2 \pi^2$ \\
$I_3=\int_0^{\infty}dr (G^2-{F^2\over 3})$\\
$I_4=\int_0^{\infty}dr r^2\bigl[ \sigma({d\over dr}+{2\over
r})\pi-\pi{d\sigma\over dr}\bigr]$\\
$I_5=\int_0^{\infty}dr r^3\pi(\pi^2/2+\tilde\sigma^2 -2\tilde\sigma
F_{\pi})$\\

\end{tabular}
\end{table}

\begin{table}
\caption{Observables for the Chiral Baryon using $m_q$=500 MeV and
$m_{\sigma}$=1200 MeV. Magnetic moments are in nuclear magnetons.
Charge radius is in fm.}
\begin{tabular}{cccccccc}
  &&This work &&&  Ref.[2]&& \\
Quantity&Quark & Meson& Total&Quark&Meson&Total&Exp.\\
\tableline
rms Charge Radius & 0.60 &&0.60&&&&0.85\\
Proton Magnetic Moment &1.39 &.28 &1.67 &1.74 & 1.13&2.88 &2.79\\

Neutron Magnetic Moment&-.98&-.28&-1.26&-1.16&-1.13&-2.29&-1.91\\

$g_A$&.98&.43&1.41&1.11&.75&1.86&1.25\\
$g_{\pi NN}{m_{\pi}\over 2 M}$&.94&.31&1.25&1.16&.379&1.53&1.0\\
Sigma commutator&&&37&&&94&$\sim$25\\
\end{tabular}
\end{table}

\begin{table}
\caption{ Energy contributions to Chiral Baryon using $m_q$=375 MeV
and $m_{\sigma}$=600 MeV. (All values are in MeV.) Fock-space soliton
solution yields A=0.878 and B=0.478.}
\begin{tabular}{lrr}
Quantity & This work &  Ref.[2]\\
\tableline
Quark eigenenergy                 & 179.4 \\
Quark kinetic energy              & 929  \\
Sigma kinetic energy               &  228 \\
Pion kinetic energy               &   146 \\
Quark-$\sigma$ interaction energy &  124 \\
Quark-pion interaction energy     &  -516 \\
Meson interaction energy          &   185 \\
Baryon Mass                       &  1096 \\
Center of Mass Correction         &  -159 \\
CM corrected Baryon Mass          &   937 \\
\end{tabular}
\end{table}

\begin{table}
\caption{Observables for the Chiral Baryon using $m_q$=375 MeV and
$m_{\sigma}$=600 MeV. Magnetic moments are in nuclear magnetons.
Charge radius is in fm.}
\begin{tabular}{cccccccc}
  &&This work && \\
Quantity&Quark & Meson& Total&Exp.\\
\tableline
rms Charge Radius & 0.82 &&0.82&0.85\\
Proton Magnetic Moment &1.72 &.46 &2.18 &2.79\\

Neutron Magnetic Moment&-1.20&-.46&-1.66&-1.91\\

$g_A$&1.19&.49&1.68&1.25\\
$g_{\pi NN}{m_{\pi}\over 2 M}$&.86&.14&1.0&1.0\\
Sigma commutator&&&86&$\sim$25\\
\end{tabular}
\end{table}
\end{document}